\documentclass[12pt]{article}
\usepackage{amssymb,graphicx}
\usepackage{epstopdf}
\usepackage{amsmath,amsfonts}
\usepackage{epsfig}
\usepackage{xcolor}
\usepackage{cite}
\usepackage{array}
\usepackage{longtable}
\usepackage[hidelinks,pdfpagelabels=false]{hyperref}

\begin{document}
\begin{flushright}
INR-TH-2020-006
\end{flushright}
\vspace{10pt}
\begin{center}
  {\LARGE \bf Toward evading the strong coupling problem \\[0.3cm] in Horndeski genesis } \\
\vspace{20pt}
Y.~Ageeva$^{a,b,c,}$\footnote[1]{{\bf email:} ageeva@inr.ac.ru}, O.~Evseev$^{a,b,}$\footnote[2]{{\bf email:} oa.evseev@physics.msu.ru}, O.~Melichev$^{d,e,}$\footnote[3]{{\bf email:} 
omeliche@sissa.it}, V.~Rubakov$^{a,b,}$\footnote[4]{{\bf email:} rubakov@ms2.inr.ac.ru}\\
\vspace{15pt}
  $^a$\textit{
Department of Particle Physics and Cosmology, Faculty of Physics,
M. V. Lomonosov Moscow State University, Vorobyovy Gory, 1-2, Moscow, 119991, Russia
  }\\
\vspace{5pt}
$^b$\textit{
Institute for Nuclear Research of
         the Russian Academy of Sciences,\\  60th October Anniversary
  Prospect, 7a, 117312 Moscow, Russia}\\
\vspace{5pt}
$^c$\textit{
Institute for Theoretical and Mathematical Physics, Leninskie Gory, GSP-1,
119991 Moscow,
Russia
}\\
\vspace{5pt}
$^d$\textit{
SISSA International School for Advanced Studies, Via Bonomea 265, 34136, Trieste, Italy
}\\
\vspace{5pt}
  $^e$\textit{
INFN, Sezione di Trieste, Trieste, Italy
  }
    \end{center}
    \vspace{5pt}


\begin{abstract}
It is of interest to understand whether or not one can construct a classical field theory description of early cosmology which would be free of initial singularity and stable throughout the whole evolution. One of the known possibilities is genesis within the Horndeski theory, which is thought to be an alternative to or a possible completion of the inflationary scenario. In this model, the strong coupling energy scale tends to zero in the asymptotic past, $t \to - \infty$, making the model potentially intractable. We point out that despite the latter property, the classical setup may be trustworthy since the energy scale of the classical evolution (the inverse of its timescale) also vanishes as $t \to -\infty$.
In the framework of a concrete model belonging to the Horndeski class,
we show that the strong coupling energy scale of the cubic interactions vastly exceeds
the classical energy scale in a certain range of parameters, 
indicating that the classical description is possible.
\end{abstract}
\vspace*{1.2cm}

\section{Introduction}

Recently, considerable attention is attracted to cosmological scenarios
without an initial classical singularity. One of them is
genesis~\cite{Creminelli:2010ba, Creminelli:2012my, Hinterbichler:2012fr, Hinterbichler:2012yn, Nishi:2015pta}
, which 
assumes that the space-time is  flat in the asymptotic past
and the energy density is initially zero. 
As time passes, 
the energy density and the Hubble rate grow, eventually reaching large values. 
This regime occurs due to the
domination of exotic matter that violates the null energy condition (NEC)
(if gravity is not described by general relativity at this stage,
 required is the violation of the null convergence condition
 \cite{Tipler:1978zz}). Later on, the energy density of exotic matter is assumed to get
 converted into the energy density of usual matter, and the conventional
 cosmological evolution sets in. This later stage may be radiation dominated,
 in which case
 genesis serves as an alternative to inflation. Another option is that
 genesis is followed by inflation; then genesis complements the
 inflationary scenario.

 A candidate class of theories with exotic matter is the
 Horndeski 
 scalar-tensor theories~\cite{Horndeski:1974wa, Fairlie:1991qe, Luty:2003vm, Nicolis:2004qq, Nicolis:2008in, Deffayet:2010qz, Kobayashi:2010cm, Padilla:2012dx} with the Lagrangian 
 containing the second derivatives of the scalar field
 and yet with the
 second-order equations of motion (for a review, see, e.g.,
 Ref.~\cite{Kobayashi:2019hrl}).
 Indeed, Horndeski theories
 admit healthy NEC-violating stages (for a review, see,
 e.g., Ref.~\cite{Rubakov:2014jja}). However,
 there is an obstacle for constructing completely
 healthy genesis models in Horndeski   theories, known as the
 ``no-go theorem''~\cite{Libanov:2016kfc,Kobayashi:2016xpl,Kolevatov:2016ppi,Akama:2017jsa}. 
 This theorem is valid in all theories of Horndeski class,
 and it is worth recalling its assumptions here. Let $h_{ij}$ and $\zeta$
 denote tensor and  scalar perturbations about a spatially flat 
 Friedmann-Lemaitre-Robertson-Walker (FLRW) classical solution with an initial genesis epoch (throughout this paper we work in the unitary gauge). 
 The unconstrained quadratic actions
 for these perturbations are
   \begin{eqnarray}
        \mathcal{ S}_{ss}=\int dt d^3x N a^3\left[
        \mathcal{ G}_S
        \frac{\dot\zeta^2}{N^2}
        -\frac{\mathcal{ F}_S}{a^2}
        \zeta_{,i}\zeta_{,i}
        \right]\label{scalar2},
    \end{eqnarray}
    \begin{eqnarray}
        \mathcal{ S}_{hh} =\int dt d^3xN\frac{a^3}{8}\left[
        \mathcal{ G}_T\frac{\dot h_{ij}^2}{N^2}-\frac{\mathcal{ F}_T}{a^2}
        h_{ij,k}h_{ij,k}\right], \label{tensor2}
    \end{eqnarray}
     where $a(t)$ is the scale factor, $N(t)$ is the lapse function and
    $\mathcal{ F}_S$, $\mathcal{ G}_S$, $\mathcal{ F}_T$, and $\mathcal{ G}_T$
    are functions of cosmic time $t$.  To avoid ghost and gradient
    instabilities, one requires that
    \begin{align}
        \mathcal{ F}_S, \mathcal{ G}_S, \mathcal{ F}_T, \mathcal{ G}_T>0.
        \label{noproblems}
    \end{align}
    The assumptions of the no-go theorem are that the background
    is nonsingular at all times, the functions
    $\mathcal{ F}_S$, $\mathcal{ G}_S$, $\mathcal{ F}_T$, and $\mathcal{ G}_T$
    do not vanish at any time and, crucially, the integral
    \begin{equation}
        \int_{-\infty}^{t} a(t) [\mathcal{ F}_T (t)+\mathcal{ F}_S(t)] dt
    \label{no-go_int}
    \end{equation}
    is divergent at the lower limit of integration. The defining property of genesis
    is $a(t) \to 1$ as $t\to - \infty$, therefore a sufficient
    condition for the latter property is that  $\mathcal{ F}_T$ and $\mathcal{F}_S$
    are finite
    as $t \to -\infty$. The no-go theorem states that under these
    assumptions, there is a gradient or ghost instability at some
    stage of the cosmological evolution; this stage may occur well
    after the initial genesis epoch.

    The no-go theorem does not hold for  theories that generalize the
    Horndeski class. This has been demonstrated explicitly
in Refs.~\cite{Cai:2016thi, Creminelli:2016zwa, Kolevatov:2017voe}
     for the
     ``beyond Horndeski'' theories~\cite{Zumalacarregui:2013pma, Gleyzes:2014dya}.
     The latter property enables one to construct completely healthy
     genesis models in beyond Horndeski
     theories~\cite{Kolevatov:2017voe,Mironov:2019qjt}.

     Here we consider another
     option. Namely, as suggested in
     Refs.~\cite{Kobayashi:2016xpl,Ijjas:2016vtq,Nishi:2016ljg},
     one works with (unextended) Horndeski theories and requires that
     the integral (\ref{no-go_int}) is convergent:
     \begin{equation}
        \int_{-\infty}^{t} a(t) [\mathcal{ F}_T (t)+\mathcal{ F}_S (t)] dt < \infty \; .
    \label{no-go_cond}
    \end{equation}
This implies that $   \mathcal{ F}_T \to 0$, $   
\mathcal{ F}_S \to 0$  as $t \to -\infty$;
as discussed in  Refs.~\cite{Kobayashi:2016xpl,Ijjas:2016vtq},
one also has  $   \mathcal{ G}_T \to 0$, $   \mathcal{ G}_S \to 0$  as $t \to -\infty$.
In this case, the
coefficients in the quadratic action for perturbations about
the classical solution tend to zero as $t \to -\infty$.
Thus, the class of models suggested in
Refs.~\cite{Kobayashi:2016xpl,Ijjas:2016vtq,Nishi:2016ljg}
is potentially problematic: not only is gravity in the asymptotic past
grossly different from general relativity, but also
the strong coupling energy scale may be expected to tend to zero
as $t\to -\infty$. By considering an explicit
model, we will see below (see also Ref.~\cite{Ageeva:2018rhc})
that this is indeed the case: the strong
coupling energy scale vanishes in the asymptotic past.\footnote{Let us
  comment on the issue of geodesic completeness. We work in the
  Jordan frame,
  so the genesis (Minkowskian) asymptotics ensure that the space-time
  is past geodesically complete for conventional matter. On the other hand,
  tensor perturbations (gravitational waves) propagate in the effective FLRW
  metric with $a^3_{eff}/N_{eff} = a^3  \mathcal{ G}_T/N$,
  $a_{eff} N_{eff} = Na  \mathcal{ F}_T$, where $a_{eff}$ and $N_{eff}$
  are effective scale factor and lapse function, respectively.
  The standard geodesic completeness condition~\cite{Borde:2001nh}
  $\int_{-\infty}^t~a_{eff}N_{eff}~dt = \infty$
  is violated in view of \eqref{no-go_cond}.
  It is not clear, however, whether or not this property, valid for
  tensor modes (gravitational waves) only, is problematic.
  Indeed, the functions  $\mathcal{ F}_T$ and $ \mathcal{ G}_T$ are
  not invariant under field redefinition; as an example, 
  for   $\mathcal{ F}_T= \mathcal{ G}_T$ and upon introducing
  canonically
  normalized field one has $a_{eff}^{(c)} = N_{eff}^{(c)} =1$, and
  the standard geodesic completeness condition holds. A similar observation
applies to the scalar sector of perturbations.}

At first sight, the latter property makes the whole construction
intractable. However, we pointed out in  Ref.~\cite{Ageeva:2018rhc}
and emphasize here that this is not necessarily the case.
Indeed, the timescale of the classical
evolution tends to infinity, and hence its inverse, the
classical energy scale, tends to zero
as $t \to -\infty$.
So, to see whether or not the classical field theory treatment is
legitimate, one has to figure out the actual strong coupling energy scale
and compare it to the inverse timescale of the classical
background evolution. The classical analysis of the background
is consistent, provided that the former energy scale much exceeds the latter.

In this paper, we consider
 genesis in a restricted class of Horndeski theories.
These genesis constructions were suggested in Ref.~\cite{Kobayashi:2016xpl},
and they obey (\ref{no-go_cond}), thus avoiding the no-go theorem.    
    To estimate the time dependence of the strong coupling scales 
    at early times, $t\to -\infty$, we study cubic terms in the action for
    perturbations $\zeta$ and $h_{ij}$ and make use of
    the naive dimensional analysis (the preliminary study of
    the scalar sector has been reported in Ref.~\cite{Ageeva:2018rhc}).
    Although our analysis is incomplete, as quartic and higher-order terms may
    yield even lower strong coupling energy scales, it does indicate that
    there may well be
    a region in the parameter space where the classical field theory
    treatment is legitimate at early times.

    This paper is organized as follows. In Sec. \ref{gen} we present the
    model and recall the properties of the classical
    solution and the quadratic action for perturbations in the asymptotic past.
    Section \ref{fullanalysis} is dedicated to
    the analysis of strong coupling resulting from the cubic
    action for perturbations. We conclude in Sec. \ref{concl}.  In
    Appendix \ref{appA}  we collect (fairly cumbersome)
    formulas
    we used in our calculations.
    While in the main text we work exclusively in the Jordan frame,
    some new insight is obtained by going to the Einstein frame; this
    aspect has to do with the peculiarities of
      strong coupling 
    in the case of a singular metric. 
    We  discuss it in 
    Appendix \ref{appB}.

\section{Preliminaries}
\label{gen}

\subsection{The model}

In this paper we follow Ref.~\cite{Kobayashi:2016xpl} and
consider
a simple subclass of the  Horndeski theories that admits stable genesis. 
The general form of the Lagrangian
for this
subclass is
    \begin{eqnarray}
    \cal L&=&G_2(\phi, X)-G_3(\phi, X)\Box \phi+ G_4(\phi)R,
    \label{Hor_L}\\
        X &=& -\frac{1}{2}g^{\mu\nu}\partial_{\mu}\phi\partial_{\nu}\phi,
    \nonumber
    \end{eqnarray}
    where $R$ is the Ricci scalar
    and $\Box \phi = g^{\mu\nu} \nabla_\mu \nabla_\nu \phi$. The metric
    signature is $(-,+,+,+)$. Unlike 
    the general Horndeski theory,
    the Lagrangian  \eqref{Hor_L} involves three arbitrary functions
    $G_{2,3,4}$ rather than four, and one of these functions, $G_4$,
    depends on $\phi$ only.

    We consider this theory at large negative times and study spatially flat
    backgrounds.
    It is convenient to use the freedom of field redefinition
    and choose the background field $\phi$ as follows:
  \begin{equation}
    e^{-\phi} = - \sqrt{2Y_0}t,
    \label{phi}
    \end{equation}  
  where $Y_0$ is a constant. Throughout this paper,  we use the unitary gauge,
  in which the field $\phi$ has the form \eqref{phi} to all orders
  of perturbation theory about homogeneous and isotropic background.\footnote{A subtle property of the Horndeski theories
  is the possible lack of strong hyperbolicity in the harmonic gauge and
  its generalizations~\cite{Papallo:2017qvl} (see, however,
  Ref.~\cite{Kovacs:2020ywu}). We think this is a peculiarity
  of generalized harmonic gauges which has to do with incomplete gauge
  fixing, whereas in the unitary gauge, weak hyperbolicity (absence of
  gradient instabilities) ensures strong hyperbolicity.
  In any case, the particular subclass  \eqref{Hor_L} 
  of the Horndeski theories is strongly hyperbolic even in the generalized
  harmonic gauge, at least in the weak
  field backgrounds~\cite{Papallo:2017qvl}.}
  We also impose the gauge in which longitudinal perturbations of
  spatial metric
  vanish and disregard vector perturbations, which are trivial. Then the
  metric, with perturbations included, is
   \begin{equation}
   \label{metric}
     ds^2=-N^2 dt^2
     +\gamma_{ij}\left( dx^i+N^i dt\right)\left(dx^j+N^j dt\right),
    \end{equation}
   where
       \begin{equation*}
        \gamma_{ij}=a^2e^{2\zeta} \left(e^{h}\right)_{ij},
    \end{equation*}
    with
    \begin{equation*}
       (e^h)_{ij}=\delta_{ij}+h_{ij}+\frac{1}{2}h_{ik}h_{kj}+\frac{1}{6}h_{ik}h_{kl}h_{lj}+\cdots
    \end{equation*}
    $\zeta$ and transverse traceless matrix
    $h= [h_{ij}]$ 
    are scalar and tensor metric perturbations, respectively, 
    while the lapse and the shift functions involve perturbations:
    \begin{equation*}
      \delta N= \alpha, \;\;\;\;\; \delta N_i=\partial_i\beta.
      \end{equation*}
    
    To make contact with Ref.~\cite{Kobayashi:2016xpl}, and 
    also for later convenience, let us rewrite the Lagrangian \eqref{Hor_L} in terms of Arnowitt-Deser-Misner
    (ADM) variables:
    \begin{align}
        \mathcal{L} =  A_2 (t, N) + A_3 (t, N) K 
        +  A_4 (K^2 - K_{ij}^2) + B_4 (t, N) R^{(3)} \text{,}
        \label{adm_lagr}
    \end{align}
    where
     $K_{ij}$ and $R^{(3)}_{ij}$
    are the extrinsic curvature and the Ricci tensor of the spatial slices,
    respectively, and we use the unitary gauge $\delta \phi = 0$.
There is a one-to-one correspondence between the
variables
$\phi$ and $X$ in the covariant Lagrangian and time variable
$t$ and the lapse function $N$ in the ADM formalism.
This correspondence involves the relation  \eqref{phi} and
    \begin{equation}
        e^{\phi}\sqrt{\frac{Y_0}{X}} = N.
        \label{gauge}
    \end{equation}
The following expressions
convert one formalism to
another~\cite{Gleyzes:2014dya, Gleyzes:2013ooa, Fasiello:2014aqa}:
    \begin{align}
        G_2 =& A_2 - 2XF_{\phi} \text{,} \label{ADM-trans2}\\
        G_3 =& - 2XF_X - F \text{,} \label{ADM-trans3}\\
        G_4 =& - A_4= B_4 \text{,} \label{ADM-trans4}
    \end{align}
where $F(\phi, X)$ is an auxiliary function, such that
    \begin{equation}
        F_X = - \frac{A_3}{\left(2X\right)^{3/2}} - \frac{B_{4\phi}}{X} \text{.}
    \end{equation}
    The subscripts $X$ and $\phi$ denote the derivatives with respect to $X$ and $\phi$, respectively.
    The equations for background are obtained by setting $N=N (t)$,
    $N^i=0$, $\gamma_{ij} = a^2(t)\delta_{ij}$ in \eqref{adm_lagr}, so that
      the action reads
    \begin{equation}
        \mathcal{S}^{(0)} = \int~dt~Na^3(A_2 +3 A_3H+6A_4H^2) \; .
    \end{equation}
      Explicitly, the equations for background are~\cite{Kobayashi:2015gga}
    \begin{subequations}
    \label{adm_eqs}
    \begin{align}
        &(NA_2)_N+3NA_{3 \, N} H+6N^2(N^{-1}A_4)_N H^2 = 0, \\
        &A_2-6A_4H^2-\frac{1}{N}\frac{d}{dt}\left( A_3+4A_4H \right)=0, 
    \end{align}
    \end{subequations}
where the Hubble parameter is $H= \dot{a}/(Na)$
and subscript $N$ denotes the derivative with respect to the
lapse function $N$.

\subsection{Getting around the no-go theorem}

In the theory \eqref{Hor_L},
the coefficients in the quadratic Lagrangian for tensor perturbations
\eqref{tensor2} are simply
    \begin{subequations}
    \label{tensor2form}
    \begin{eqnarray}
&\mathcal{F}_T = 2G_4 = 2B_4 \; , \\
        &\mathcal{G}_T = 2G_4 = 2B_4 \; .
    \end{eqnarray}
    \end{subequations}
Therefore, the necessary condition \eqref{no-go_cond} for evading the no-go theorem [together
with the genesis condition
$a(t) \to 1$ as $t \to -\infty$]
means that $G_4 (\phi)$ sufficiently rapidly
tends to zero as $t\to -\infty$. The other two Lagrangian functions
in \eqref{Hor_L} are to be chosen in such a way that the background solution to
the field equations describes genesis, i.e.,  $a\to 1$, $N \to 1$
as $t\to-\infty$.
Note that the requirement that $G_4 \to 0$ as $t \to -\infty$ immediately
implies that the strong coupling energy scale tends to zero in the
asymptotic past: $G_4$ serves as an effective Planck mass squared.
 
An explicit construction is conveniently described in the ADM language.
An example that we study in this paper is
given in Ref.~\cite{Kobayashi:2016xpl}: 
    \begin{subequations}
    \label{adm_func_lagr}
    \begin{align}
        &A_2 =  f^{-2\alpha -2 -\delta} a_2 (N) \text{,} \\ 
        &A_3 =  f^{-2\alpha -1 -\delta} a_3 (N) \text{,} \\
        & B_4 = - A_4 = f^{-2\alpha} \text{,}
        \label{B_4}
    \end{align}
    \end{subequations}
where $\alpha$ and $\delta$ are constant parameters satisfying
    \begin{equation}
        2\alpha > 1 + \delta \; ,  \hspace{5mm}  \delta > 0 \; , \label{Kob_cond}  
    \end{equation}
and $f(t)$ is some function of time, which has the following asymptotics
as $t \rightarrow - \infty $:
    \begin{equation}
        f \approx -ct,  \hspace{5mm} c=\text{const}>0 \text{.}
    \end{equation}
    The functions $a_2$ and $a_3$ entering
    \eqref{adm_func_lagr} are given by
    \begin{align}
        &a_2(N) = -\frac{1}{N^2} + \frac{1}{3N^4} \text{,}\\
        &a_3(N) = \frac{1}{4N^3}\text{.}
    \label{smol_a}
    \end{align}
The solution to \eqref{adm_eqs} has the following  asymptotics
at early times, $t \rightarrow - \infty$:
    \begin{equation}
        H \approx \frac{\chi}{(-t)^{1+\delta}} \; ,
    \label{hub}
    \end{equation}
    \begin{equation}
        a \approx 1 + \frac{\chi}{\delta (-t)^\delta}, \quad N \approx 1\; ,
    \label{lapse_scale}
    \end{equation}
where $\chi$ is the combination of the Lagrangian parameters:
    \begin{equation}
        \chi = \frac{\frac{2}{3}+
        \frac{c}{4}\left(2\alpha+1+\delta\right)}{4(2\alpha
        +1+\delta)c^{2+\delta}} \; .
    \end{equation}
Thus, the background indeed describes the genesis stage at early times.

The purpose of this paper is to see whether the classical treatment
of this stage is legitimate.
To this end, we make use of the naive dimensional analysis
and find the early-time asymptotics of the strong coupling energy
scales dictated by various cubic (and also quadratic)
terms 
in the Lagrangian for
perturbations. These scales have inverse power-law behavior in $t$. 
We compare these scales with the energy scale
characteristic of the classical evolution. The latter equals the
inverse timescale
    \begin{equation}
        E_{class} \propto  \frac{\dot{H}}{H} \propto (-t)^{-1}
        \label{classicscale}
    \end{equation}
[another classical energy scale $H$ is lower; see Eq.~\eqref{hub}].
Thus, if the strong coupling energy scales decrease slower than $(-t)^{-1}$
as
$t \to -\infty$, the classical treatment of the background evolution
is legitimate,  assuming that interactions of higher
than third order do not induce lower energy scales than cubic ones.

\subsection{Quadratic actions for perturbations}

The quadratic action for tensor perturbations is given by Eq.~\eqref{tensor2},
where the explicit form of  $\mathcal{ F}_T$ and $ \mathcal{ G}_T$
is determined by \eqref{tensor2form} and \eqref{B_4}. The early-time
asymptotics are
    \begin{equation}
        \mathcal{ F}_T \propto (-t)^{-2\alpha},\;\;\;\;\;\;\;
        \mathcal{ G}_T\propto (-t)^{-2\alpha} \; , \;\;\;\; \text{as} \;\;\;\;
        t \to -\infty \;.
\label{tensor2asymp}
    \end{equation}
The general expressions for the coefficients
$\mathcal{ F}_S$ and $ \mathcal{ G}_S$ in the action for scalar perturbations
are given in  Appendix \ref{appA} [Eq.~\eqref{scalar2formAPP}].
They lead to the following early-time asymptotics:
    \begin{equation}
        \mathcal{ F}_S \propto  (-t)^{-2\alpha+\delta},\;\;\;\;\;\;\;
        \mathcal{ G}_S\propto (-t)^{-2\alpha+\delta}
        \; , \;\;\;\; \text{as} \;\;\;\;
        t \to -\infty \;.
    \label{scalar2asymp}
    \end{equation}
In view of \eqref{Kob_cond}, \eqref{tensor2asymp} and \eqref{scalar2asymp}, the integral \eqref{no-go_cond} 
is convergent indeed. The price to pay is that $\mathcal{ F}_T$, $\mathcal{ G}_T$, $\mathcal{ F}_S$, and
$\mathcal{ G}_S$ vanish in the asymptotic past,
which may signalize the strong coupling problem coming from either a
scalar, tensor, or mixed scalar-tensor sector.

\subsection{Strong coupling scales from quadratic action}
The simplest estimates for strong coupling scales are obtained by
considering interactions of external sources via exchange of either
tensor or scalar mode. In the former case, the effective ``Planck mass''
is of order  $\mathcal{ F}_T^{1/2} \sim  \mathcal{ G}_T^{1/2} \propto
(-t)^{-\alpha}$. This strong coupling energy scale is higher than
the classical scale $(-t)^{-1}$ [Eq.~\eqref{classicscale}], provided that
\begin{equation*}
\alpha < 1 .
\end{equation*}
We will see in Sec.~\ref{subsec:tens} that the same bound follows from the
analysis of cubic interactions of tensor modes.

A bold use of the same argument for scalar mode exchange
would yield the strong coupling scale
$\mathcal{ F}_S^{1/2} \sim  \mathcal{ G}_S^{1/2} \propto
(-t)^{-\alpha + \delta/2}$, which would not give anything new.
Let us be not so bold, however. We recall that in Horndeski theory and in
the absence of extra background matter,
the gravitational interaction of static sources \footnote{To evaluate the interaction between two sources one should move away from the unitary gauge and rewrite the action in the Newtonian gauge, explicitly reintroducing the scalar field fluctuations; see App. B in Ref. \cite{Cusin:2017mzw} for details.} at distances shorter than
the evolution scale $t$ is characterized by effective gravitational
constant~\cite{DeFelice:2011hq}
\begin{equation*}
G_{eff} = \frac{(\dot{\Theta} + H \Theta) \mathcal{ F}_S
  +(\dot{\mathcal{ G}}_T - \dot{\Theta} \mathcal{ G}_T/\Theta)^2}{\Theta^2
  \mathcal{ F}_S} \; ,
\end{equation*}
where the parameter $\Theta$ is defined in Appendix \ref{appA}
[Eq.~\eqref{defTheta}] and has asymptotic behavior
$\Theta \propto  (-t)^{-2\alpha-\delta-1}$ [see Eq.~\eqref{asyTheta}].
We find that the energy scale associated with interaction of
nonrelativistic sources is actually of order
$G_{eff}^{-1/2} \propto (-t)^{-\alpha - \delta/2}$. We again require
that it is higher than $(-t)^{-1}$ and obtain a stronger bound
\begin{equation}
\alpha + \frac{\delta}{2} < 1.
\label{Gbound}
\end{equation}
We will see in Sec.~\ref{sec_3_scal},
however, that the bound obtained by considering cubic
interactions of scalar modes is stronger than \eqref{Gbound}
[see Eq.~\eqref{NSC_scalars}], so quadratic action alone is
insufficient to figure out the range of parameters where the
classical treatment is legitimate.
\section{Third-order analysis}
\label{fullanalysis}
We consider interaction terms in various sectors separately.
\subsection{Scalar sector}
\label{sec_3_scal}
We begin with the cubic action involving scalar perturbations only
and use the results obtained in
Refs.~\cite{Gao:2011qe, DeFelice:2011uc, Gao:2012ib}.
We calculate the unconstrained
action in Appendix A. It has the following
form:
    \begin{align}
        \mathcal{S}^{(3)}_{\zeta\zeta\zeta}= &\int dt d^3x
        \sum_{i=1}^{17} \mathbb{L}^{(i)}_{\zeta \zeta \zeta}
        =  \int Ndt\text{ }a^3d^3x \left\{ \Lambda_1 \frac{\dot{\zeta}^3}{N^3} 
        + \Lambda_2 \frac{\dot{\zeta}^2}{N^2}\zeta 
        + \Lambda_3 \frac{\dot{\zeta}^2}{N^2} \partial^2 \zeta  
        +\Lambda_4 \frac{\dot{\zeta}}{N}\zeta \partial^2 \zeta\right. \nonumber \\
        &  + \Lambda_5 \frac{\dot{\zeta}}{N} \left(\partial_i \zeta \right)^2 
        +\Lambda_6 \zeta \left(\partial_i \zeta \right)^2 
        + \Lambda_7 \frac{\dot{\zeta}}{N} \left(\partial^2 \zeta \right)^2 
        + \Lambda_8 \zeta \left(\partial^2 \zeta \right)^2 
        + \Lambda_9 \partial^2 \zeta \left(\partial_i \zeta \right)^2 
        \nonumber \\
        &+ \Lambda_{10} \frac{\dot{\zeta}}{N} \left(\partial_i \partial_j \zeta \right)^2 
        + \Lambda_{11} \zeta \left(\partial_i \partial_j \zeta \right)^2 
        + \Lambda_{12} \frac{\dot{\zeta}}{N} \partial_i \zeta \partial_i \psi 
        + \Lambda_{13} \partial^2 \zeta \partial_i \zeta \partial_i \psi 
        +  \Lambda_{14} \frac{\dot{\zeta}}{N} \left(\partial_i \partial_j \psi \right)^2 
        \nonumber \\
        &+ \Lambda_{15} \zeta \left(\partial_i \partial_j \psi \right)^2
        + \left.
        \Lambda_{16} \frac{\dot{\zeta}}{N} \partial_i \partial_j \zeta \partial_i \partial_j \psi
        + \Lambda_{17} \zeta \partial_i \partial_j \zeta \partial_i \partial_j \psi \right\}, \label{S3scalar}
    \end{align}
where $\partial^2 = \partial_i \partial_i$,
    \begin{equation}
        \psi=(1/N)\partial^{-2}\dot\zeta
        \nonumber
    \end{equation}
and  $\Lambda_1, \dots , \Lambda_{17}$ are
functions of time $t$.  All of them have power-law behavior at early times $t\to-\infty$:
    \begin{equation}
        \Lambda_i \propto (-t)^{x_i},
    \end{equation}
where $x_i$ are combinations of the parameters
    $\alpha$ and $\delta$.
The general formulas for the coefficients
 $\Lambda_1, \dots , \Lambda_{17}$
are collected in Appendix \ref{appA}. 

There is a subtlety regarding the form of the third-order action \eqref{S3scalar}.
It has been claimed 
in Ref.\cite{Gao:2012ib} that the cubic action for scalar perturbation $\zeta$ 
containing only five different terms
(rather than 17) can be obtained by integration by parts. We find it more straightforward to 
work with all 17 terms. Thus, the conditions for the validity of the classical treatment 
of the early genesis, which we are about to derive, are sufficient but, generally speaking, 
not necessary: the reduction to five terms may in principle lead to cancellations 
and to increase of the strong coupling energy scale as compared to our analysis.

The naive dimensional analysis proceeds as follows
(see also Ref.~\cite{Ageeva:2018rhc}).
For power-counting purposes,
every term $\mathbb{L}^{(i)}_{\zeta \zeta \zeta}$ in the cubic Lagrangian ($i = \overline{1,17}$) 
is schematically written as 
    \begin{equation}
        \label{term_lagr}
        \mathbb{L}^{(i)}_{\zeta \zeta \zeta} \propto \Lambda_i\cdot  \zeta^3 \cdot (\partial_t)^{a_i} \cdot
        ( \partial )^{b_i}
        \; ,
    \end{equation}
where $a_i$ and $b_i$ are the
numbers of temporal and spatial derivatives, respectively.
One introduces the canonically
normalized field $\pi$ instead of $\zeta$. Since $a(t)$ and $N(t)$
tend to constants as  $t \to -\infty$, and $\mathcal{F}_S \propto
\mathcal{G}_S$, we have
(modulo a time-independent factor)
    \begin{equation}
        \pi = \sqrt{2 \mathcal{G}_S} \zeta \propto (-t)^{-\alpha +\delta/2} \zeta 
        \; , 
        \label{scalarcanonic}
    \end{equation}
    The fact that the coefficient here tends to zero as  $t \to -\infty$
    is crucial for what follows.
In terms of the canonically normalized field $\pi$ one rewrites (\ref{term_lagr}) as \footnote{Since $\zeta \propto (-t)^{\alpha - \delta/2} \pi $,
      we have $\dot{\zeta} \propto  (-t)^{\alpha - \delta/2} (\dot{\pi}
      +\mbox{const}\cdot \pi/t)$. The second term here generates additional
    vertices in the interaction Lagrangian written in terms of $\pi$. However,
    we are interested in energies exceeding $(-t)^{-1}$, for which
    $\dot{\pi} \gg \pi/t$, so these additional vertices are negligible.
    In other words, along with the Lagrangian \eqref{a_fin} there is an
    interaction term  $t^{-1} \cdot \hat{\Lambda}_i\cdot
    \pi^3 \cdot (\partial_t)^{a_i-1} \cdot
    ( \partial )^{b_i}$, and it is straightforward to check that
    the strong coupling scale associated
    with the latter term is higher than the scale \eqref{strong-3scalar}
    inferred from the Lagrangian  \eqref{a_fin}, provided that the scale
     \eqref{strong-3scalar} is higher than the classical scale $(-t)^{-1}$.
  So, it is sufficient to consider the Lagrangian \eqref{a_fin} only.}
    \begin{equation}
        \label{a_fin}
        \mathbb{L}^{(i)}_{\zeta \zeta \zeta} \propto \hat{\Lambda}_i\cdot  \pi^3 \cdot (\partial_t)^{a_i} \cdot
        ( \partial )^{b_i} \; , 
    \end{equation}
where
    \begin{equation}
        \hat{\Lambda}_i = \Lambda_i\mathcal{G}_S^{-3/2} =\Lambda_i (-t)^{-\frac{3}{2}(\delta - 2\alpha)}
        \propto  (-t)^{x_i-\frac{3}{2}(\delta - 2\alpha)} \; .
    \end{equation}
Now, the dimension of $\hat{\Lambda}_i$ is $1-a_i-b_i$, so the strong 
coupling energy scale associated with the term
$\mathbb{L}^{(i)}_{\zeta\zeta\zeta}$ is
    \begin{equation}
    \label{strong-3scalar}
        E^{\zeta\zeta\zeta,(i)}_{strong} \propto \hat{\Lambda}_i^{- \frac{1}{a_i + b_i -1}}
        \propto (-t)^{-\frac{x_i + 3\alpha -     3\delta/2}{a_i + b_i -1}} \; .
    \end{equation}
By requiring that  $ E_{class} \ll  E^{\zeta\zeta\zeta,(i)}_{strong}$, where
$E_{class}$ is
the energy scale of the classical evolution
\eqref{classicscale},
we find
the condition for the legitimacy of the classical treatment
of the early evolution:
    \begin{equation}
        x_i+3\alpha-\frac{3}{2}\delta < a_i+b_i-1 , \;\;\;\;\;
        \mbox{for~all} \;\;\;
        i = \overline{1,17}\; .
    \label{NSC_general_scalar}
    \end{equation}
We collect the properties of the 17 terms in the cubic Lagrangian
\eqref{S3scalar}, as well as the resulting constraints
on $\alpha$ and $\delta$ coming from \eqref{NSC_general_scalar}, in
Table~\ref{scalarcondtab}.

By inspecting this table we find that all constraints \eqref{NSC_general_scalar}
are satisfied provided that
    \begin{equation}
        0 < \delta < \frac{1}{4}, 
        \qquad 2 - 3\delta > 2\alpha > 1 + \delta \; ,
    \label{NSC_scalars}
    \end{equation}
where we also recall \eqref{Kob_cond}. 
This completes the analysis of the scalar sector.
\newpage

\renewcommand{\arraystretch}{1.4}

\setlength{\LTcapwidth}{\textwidth}
\begin{longtable}{cccccr}
\hline
\hline
Term & $\big[\hat{\Lambda}_i\big]$ & $x_i$ & $a_i$ & $b_i$ & Condition 
\\
\hline
\hline
\endhead
$\Lambda_1 \big( \dot{\zeta}/N  \big)^3$                                           
& $ - 2$ & $1 - 2\alpha + 3\delta$ & 3 & 0 & $2 \alpha + 3 \delta < 2$\\
\hline
$\Lambda_2 \big( \dot{\zeta}/N \big)^2 \zeta$                                    
& $ - 1$ & $- 2\alpha + 2\delta$   & 2 & 0 & $2 \alpha + \delta < 2$\\
\hline
$\Lambda_3 \big( \dot{\zeta}/N \big)^2 \partial^2 \zeta$                       
& $ - 3$ & $2 - 2\alpha + 3\delta$ & 2 & 2 & $2 \alpha + 3 \delta < 2$\\
\hline
$\Lambda_4 (\dot{\zeta}/N) \zeta  \partial^2 \zeta$                              
& $ - 2$ & $1 - 2\alpha + 2\delta$ & 1 & 2 & $2 \alpha + \delta < 2$\\
\hline
$\Lambda_5 (\dot{\zeta}/N) \big( \partial_i \zeta \big)^2$                            & $ - 2$ & $1 - 2\alpha + 2\delta$  & 1 & 2 & $2 \alpha + \delta < 2$\\
\hline
$\Lambda_6 \zeta \big( \partial_i \zeta \big)^2$                              & $ - 1$ & $- 2\alpha$             & 0 & 2 & $2 \alpha - 3 \delta < 2$\\
\hline
$\Lambda_7 (\dot{\zeta}/N) \big( \partial^2 \zeta \big)^2$                            & $ - 4$ & $3 - 2\alpha + 3\delta$  & 1 & 4 & $2 \alpha + 3\delta < 2$\\
\hline
$\Lambda_8 \zeta \big( \partial^2 \zeta \big)^2$                              & $ - 3$ & $2 - 2\alpha + 2\delta$  & 0 & 4 & $2 \alpha + \delta < 2$\\
\hline
$\Lambda_9 \partial^2 \zeta \big( \partial_i \zeta \big)^2$                   & $ - 3$ & $2 - 2\alpha + 2\delta$  & 0 & 4 & $2 \alpha + \delta < 2$\\
\hline
$\Lambda_{10} (\dot{\zeta}/N) \big( \partial_i \partial_j \zeta \big)^2$              & $ - 4$ & $3 - 2\alpha + 3\delta$  & 1 & 4 & $2 \alpha + 3\delta < 2$\\
\hline
$\Lambda_{11} \zeta \big( \partial_i \partial_j \zeta \big)^2$                & $ - 3$ & $2 - 2\alpha + 2\delta$  & 0 & 4 & $2 \alpha + \delta < 2$\\
\hline
$\Lambda_{12} (\dot{\zeta}/N) \partial_i \zeta \partial^i \psi$                       & $ - 1$ & $ - 2\alpha + 2\delta$   & 2 & 0 & $2 \alpha + \delta < 2$ \\
\hline
$\Lambda_{13} \partial^2 \zeta \partial_i \zeta \partial^i \psi$              & $ - 2$ & $1 - 2\alpha + 2\delta$ & 1 & 2 & $2 \alpha + \delta < 2$ \\
\hline
$\Lambda_{14}  (\dot{\zeta}/N) \big( \partial_i \partial_j \psi \big)^2$    & $ - 2$ & $1 - 2\alpha + 3\delta$ & 3 & 0 & $2 \alpha + 3 \delta < 2$ \\
\hline
$\Lambda_{15} \zeta \big( \partial_i \partial_j \psi \big)^2$      & $ - 1$ & $- 2\alpha + 2\delta$   & 2 & 0 & $2 \alpha + \delta < 2$ \\
\hline
$\Lambda_{16} (\dot{\zeta}/N) \partial_i \partial_j \zeta \partial^i \partial^j \psi$ & $ - 3$ & $2 - 2\alpha + 3\delta$ & 2 & 2 & $2 \alpha + 3 \delta < 2$ \\
\hline
$\Lambda_{17} \zeta \partial_i \partial_j \zeta \partial^i \partial^j \psi$   & $ - 2$ & $1 - 2\alpha + 2\delta$ & 1 & 2 & $2 \alpha + \delta < 2$ \\
\hline
\hline\\
\caption{Strong coupling analysis for the scalar sector of metric
  perturbations. The columns are: (i) the terms in the cubic Lagrangian;
  (ii) the dimension of the coefficient $\hat{\Lambda}_i$;
  (iii) the exponent of the asymptotic behavior $x_i$;
  (iv) the number of temporal derivatives $a_i$;
  (v) the number of spatial derivatives $b_i$; and
  (vi) the condition for the absence of strong coupling obtained from
  \eqref{NSC_general_scalar}.}
\label{scalarcondtab}
\end{longtable}

\renewcommand{\arraystretch}{1}

\subsection{Scalar-tensor-tensor sector}
The cubic interaction terms involving two tensors and one
scalar are~\cite{Gao:2012ib}
    \begin{eqnarray}
        \mathcal{ L}_{\zeta hh}&=&a^3\left[
        d_1\zeta\frac{\dot h_{ij}^2}{N^2}
        +\frac{d_2}{a^2}\zeta h_{ij,k}h_{ij,k}
        +d_3\psi_{,k}\frac{\dot h_{ij}}{N}h_{ij,k}
        +d_4\frac{\dot\zeta}{N}\frac{\dot h_{ij}^2}{N^2}
        +\frac{d_5}{a^2}\partial^2\zeta \frac{\dot h_{ij}^2}{N^2}\right.
        \nonumber\\&&
        \left.
        +d_6\psi_{,ij}\frac{\dot h_{ik}}{N}\frac{\dot h_{jk}}{N}
        +\frac{d_7}{a^2}\zeta_{,ij}\frac{\dot h_{ik}}{N}\frac{\dot h_{jk}}{N}\right].
        \label{lagr_stt}
    \end{eqnarray}
General expressions for coefficients $d_i$ are collected
in Appendix \ref{appA}; in our particular class of models \eqref{Hor_L}
we have
    \begin{equation*}
        d_4=d_5= d_6 = d_7=0.
    \end{equation*}
All $d_i$ have again power-law asymptotic behavior:
    \begin{equation}
        d_i \propto (-t)^{y_i},
    \end{equation}
with $y_i$ being combinations of $\alpha$ and $\delta$.
The structure of the terms in the Lagrangian is
    \begin{equation}
        \label{term_lagr_shh}
        \mathbb{L}^{(i)}_{\zeta h h} \propto d_i\cdot  \zeta h^2 \cdot (\partial_t)^{a_i} \cdot
        ( \partial )^{b_i}
        \; ,
    \end{equation}
where $h$ is the notation for tensor perturbation.
We introduce canonically normalized scalar perturbation $\pi$
via  (\ref{scalarcanonic}) and also
canonically normalized tensor perturbation (recall that $\mathcal{F}_T = \mathcal{G}_T$)
    \begin{equation}
        q_{ij} = \sqrt{2\mathcal{G}_T} h_{ij} \propto (-t)^{-\alpha} h_{ij}
        \; . 
        \label{tensorcanonic}
    \end{equation}
In terms of 
the canonically normalized fields, the terms in the Lagrangian read
    \begin{equation}
        \mathbb{L}^{(i)}_{\zeta h h} \propto \hat{d}_i\cdot  \pi q^2 \cdot (\partial_t)^{a_i} \cdot
        ( \partial )^{b_i} , 
        \label{a_fin_shh}
    \end{equation}
where
    \begin{equation}
        \hat{d}_i \propto d_i\mathcal{G}_S^{-1/2}\mathcal{G}_T^{-1}
        \propto  (-t)^{y_i-\left(\frac{\delta}{2} - 3\alpha\right)} \; .
    \end{equation}
Therefore, the strong coupling energy scale is
    \begin{equation}
        E^{\zeta h h,(i)}_{strong} \propto \hat{d}_i^{- \frac{1}{a_i + b_i -1}}
        \propto (-t)^{-\frac{y_i-\left(\frac{\delta}{2} - 3\alpha\right)}{a_i + b_i -1}} \; .
    \end{equation}
The requirement that $E_{class} \ll   E^{\zeta h h,(i)}_{strong}$ gives
    \begin{equation}
        y_i+3\alpha-\frac{\delta}{2} < a_i+b_i-1.
    \label{NSC_stt}
    \end{equation}
The properties of the terms in the Lagrangian \eqref{lagr_stt} as well as
the explicit forms of inequality \eqref{NSC_stt} are given in
Table~\ref{crosscondtab}. We find that the condition \eqref{NSC_stt}
is weaker than the bound \eqref{NSC_scalars} obtained by considering the
scalar sector.
    
\subsection{Scalar-scalar-tensor sector}
In the one tensor and two scalar case, the cubic action is written
as~\cite{Gao:2012ib}
    \begin{eqnarray}
        \mathcal{ L}_{\zeta \zeta h}&=&a^3\left[\frac{c_1}{a^2}
        h_{ij}\zeta_{,i}\zeta_{,j}
        +\frac{c_2}{a^2}\frac{\dot h_{ij}}{N}\zeta_{,i}\zeta_{,j}
        +c_3\frac{\dot h_{ij}}{N}\zeta_{,i}\psi_{,j}
        +\frac{c_4}{a^2}\partial^2h_{ij}\zeta_{,i}\psi_{,j}\right.
        \nonumber\\&&
        \left.
        +\frac{c_5}{a^4}\partial^2 h_{ij}\zeta_{,i}\zeta_{,j}
        +c_6\partial^2 h_{ij}\psi_{,i}\psi_{,j}\right]. \label{ssh}
    \end{eqnarray}
All coefficients are given in Appendix \ref{appA} and
in our particular class of models \eqref{Hor_L}
we have
    \begin{equation*}
        c_4=c_5=0 .
    \end{equation*}
The coefficients again have power-law behavior,
$c_i \propto (-t)^{z_i}$, where
$z_i$ are combinations of $\alpha$ and $\delta$.
The terms in the Lagrangian have the following form:
$\mathbb{L}^{(i)}_{\zeta \zeta h} \propto c_i\cdot  \zeta^2 h \cdot (\partial_t)^{a_i} \cdot
( \partial )^{b_i} $.
We apply the same procedure as before, express the Lagrangian in terms
of canonically normalized fields,
$\mathbb{L}^{(i)}_{\zeta \zeta h} \propto \hat{c}_i\cdot
\pi^2 q \cdot (\partial_t)^{a_i} \cdot
( \partial )^{b_i} $
with
    \begin{equation}
        \hat{c}_i \propto c_i\mathcal{G}_S^{-1}\mathcal{G}_T^{-1/2}
        \propto  (-t)^{z_i-\left(\delta - 3\alpha\right)}
    \end{equation}
and find that the condition
$E_{class} \ll   E^{\zeta \zeta h,(i)}_{strong}$ is equivalent to
    \begin{equation}
        z_i+3\alpha-\delta < a_i+b_i-1.
    \end{equation}
The results are summarized in Table~\ref{crosscondtab}. We again
find that the bounds are weaker than in the scalar case.
\newpage
\setlength{\LTcapwidth}{\textwidth}
\renewcommand{\arraystretch}{1.4}

\begin{longtable}{cccccr}
\hline
\hline
Term & $[\hat{d}_i]([\hat{c}_i] \text{ for } \mathcal{ L}_{\zeta \zeta h})$ & $y_i (z_i \text{ for } \mathcal{ L}_{\zeta \zeta h})$ & $a_i$ & $b_i$ & Condition \\
\hline
\hline
\multicolumn{6}{c}{Two tensors and one scalar}\\
\hline
\hline
$d_1 \zeta(\dot{h}_{ij}/N)^2  $                                             & $ - 1$ & $- 2\alpha + \delta$ & 2 & 0 & $2\alpha +  \delta < 2$\\
\hline
$d_2 \zeta h_{ij,k} h_{ij,k}$                                       & $ - 1$ & $- 2\alpha + \delta$   & 0 & 2 & $2\alpha +  \delta < 2$\\
\hline
$d_3 \psi_{,k} (\dot{h}_{ij}/N)h_{ij,k}$                            & $ - 1$ & $- 2\alpha + \delta$ & 2 & 0 & $2\alpha +  \delta < 2$\\
\hline
\hline
\hline
\multicolumn{6}{c}{Two scalars and one tensor}\\
\hline
\hline
$c_1 h_{ij} \zeta_{,i} \zeta_{,j} $                            & $ - 1$ & $- 2\alpha + \delta$  & 0 & 2 & $ \alpha < 1$\\
\hline
$c_2 (\dot{h}_{ij}/N) \zeta_{,i} \zeta_{,j}$                              & $ - 2$ & $1 - 2\alpha + 2\delta$             & 1 & 2 & $ \alpha + \delta < 1$\\
\hline
$c_3 (\dot{h}_{ij}/N) \zeta_{,i} \psi_{,j}$                            & $ - 1$ & $-2\alpha + 2\delta$  & 2 & 0 & $ \alpha + \delta < 1$\\
\hline
\hline
$c_6 \partial^2 h_{ij} \psi_{,i} \psi_{,j}$              & $ - 1$ & $- 2\alpha + 2\delta$  & 2 & 0 & $ \alpha + \delta < 1$\\
\hline
\hline\\
\caption{Strong coupling analysis for the mixed sectors
  of metric perturbations.
  The columns are the same as in  Table \ref{scalarcondtab}.}
\label{crosscondtab}
\end{longtable}
\renewcommand{\arraystretch}{1}

\subsection{Tensor sector}
\label{subsec:tens}
The Lagrangian involving three tensors was derived in Ref.~\cite{Gao:2011vs}:
    \begin{eqnarray}
        \mathcal{ L}_{hhh}=
        a^3\left[
        \frac{\mu }{12N^3} \dot h_{ij}\dot h_{jk}\dot h_{ki}
        +\frac{\mathcal{ F}_T}{4a^2}\left(h_{ik}h_{jl}
        -\frac{1}{2}h_{ij}h_{kl}\right)h_{ij,kl}
        \right]\; ,
        \label{threetensor}
    \end{eqnarray}
    where   $\mu =\dot\phi XG_{5X}$. In our class of models \eqref{Hor_L}
    we have
    $\mu =0$, so there is
    only one term to analyze. With $\mathcal{ F}_T\propto (-t)^{-2\alpha}$, the
Lagrangian expressed through 
    the canonically normalized field is
    \begin{equation}
        \label{a_fin_hhh}
        \mathbb{ L}_{hhh} \propto \mathcal{ F}_T\cdot (2\mathcal{G}_T)^{-3/2} 
        \cdot  q^3 \cdot (\partial_t)^{a_T} \cdot
        ( \partial )^{b_T}.  
    \end{equation}
In the three tensor case we have the following strong coupling energy scale: 
    \begin{equation}
      E^{hhh,(T)}_{strong} \propto
      (\mathcal{ F}_T\cdot (2\mathcal{G}_T)^{-3/2})^{- \frac{1}{a_T + b_T -1}}
        \propto (-t)^{-\frac{\alpha}{a_T + b_T -1}} \;,
    \end{equation}
    where $a_T = 0$ and $b_T = 2$ in view of (\ref{threetensor}).
  Therefore, the requirement $E_{class} \ll  E^{hhh, (T)}_{strong}$ gives
    \begin{equation}
      \alpha < 1.
    \end{equation}
    Of course, this bound could have been  obtained directly by inspection
    of the last term in the original Lagrangian \eqref{Hor_L}: the coefficient
    $G_4$ serves as the effective Planck mass squared, so the
    strong coupling scale is $G_4^{1/2} \propto \mathcal{G}_T^{1/2}
    \propto (-t)^{-\alpha}$. Other terms in \eqref{Hor_L} do not contain cubic
    self-interactions of transverse traceless $h_{ij}$.

\begin{figure}[h]
\centering
\includegraphics[width=14cm]{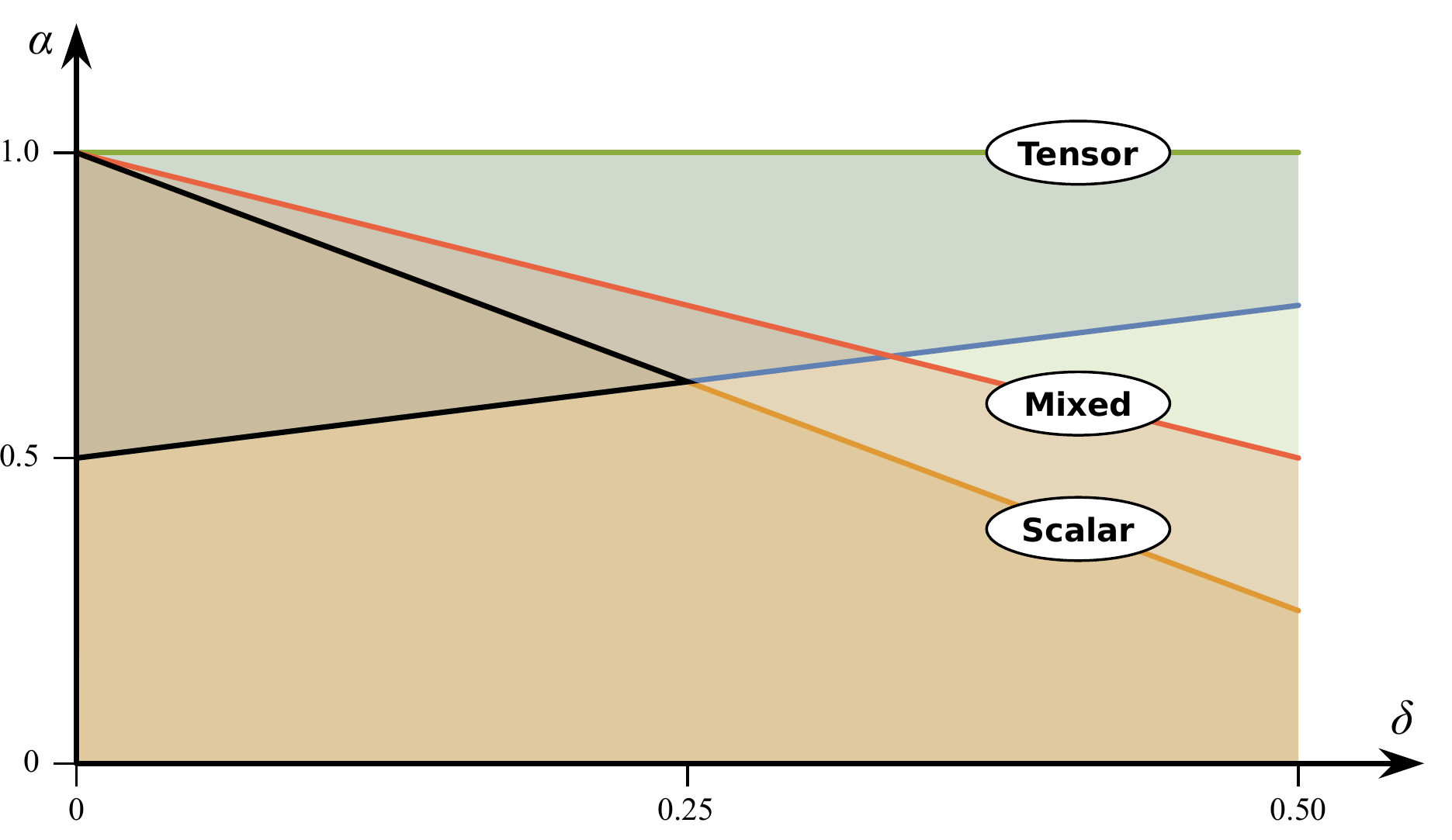}
\caption{The space of parameters $\alpha$ and $\delta$
  determining the Lagrangian functions
  \eqref{adm_func_lagr}. 
  The region above the blue line
  corresponds to the parameters which yield the Minkowski
  spacetime as $t\to-\infty$ and evade the no-go theorem of
  Ref.~\cite{Kobayashi:2016xpl}. The area
  below the  green line ($\alpha<1$) is the
  region free of  the strong coupling problem inferred from
  the tensor sector analysis at the cubic level. Similarly,
  regions below the
  red ($\alpha<1-\delta$) and orange
  ($2\alpha<2-3\delta$) lines are obtained by
  the analysis of
  the mixed
  and scalar sectors, respectively. 
  The black-framed area
  shows the allowed range of the
  parameters $\alpha$ and $\delta$ free of the strong coupling
  problem inferred from the analysis of all cubic 
  interactions and free of instabilities.}
\label{possible_param}
\end{figure}

    We summarize our results in 
    Fig.~\ref{possible_param}. The black framed triangle in this figure
    shows the region \eqref{NSC_scalars} where the early-time
    asymptotics of
    the classical energy scale
    is lower than the asymptotics of
    the  strong coupling scales found by studying all cubic
    interactions between the perturbations. As
      we discussed in Sec.\ref{sec_3_scal}, the actual region of the parameter 
    space, where the latter property holds, may be larger than our triangle. 

\section{Conclusion}
\label{concl}

We have studied the nonsingular genesis scenario
in the framework of the Horndeski theory, which is capable
of avoiding the gradient instability at the expense of potential
strong coupling problem. 
The model of genesis presented in Ref.~\cite{Kobayashi:2016xpl}
has been used as an example that gives an explicit asymptotic solution
at early times ($t\to-\infty$). 
The parameters $\alpha$ and $\delta$ that determine the
Lagrangian of  this model were chosen in the range consistent with
 stable genesis, i.e.,
violation of the no-go theorem  \cite{Kobayashi:2016xpl}. 

We have seen that with the additional restrictions on 
$\alpha$ and $\delta$, the classical genesis 
solution is away from the strong coupling regime inferred from the 
study of all cubic interactions of metric perturbations: scalar, mixed and tensor.
These restrictions came  from the requirement that all characteristic 
energy scales, associated with cubic action,  must be larger than
the classical energy scale (in our model it is
$E_{class}\propto\dot{H}/H$). 
This opens up the possibility that the Universe starts up with
very low quantum gravity energy scale (the effective Planck mass
asymptotically vanishes as $t \to -\infty$), and yet its classical
evolution is so slow that the classical field theory description
remains valid. We presented the healthy region of parameters $\alpha$ 
and $\delta$ in Fig.\ref{possible_param}. 

In Appendix \ref{appB} we also noted a non-trivial point concerning the strong 
coupling regime in models with singular metric.
Namely, in that case,
healthy behavior of the coefficients in the quadratic action
does not necessarily imply that the classical treatment of the background
evolution is legitimate.

Even though our analysis has led to
a promising outcome, it is
certainly incomplete, as there is no guarantee that the fourth
and higher order
interactions will give strong coupling energy
scales higher than or equal to the ones we have found by studying the
cubic interactions. We plan to turn to this issue in the future.

\section*{Acknowledgments} 
We are indebted  to   E.~Babichev, T.~Kobayashi, S.~Mironov, 
A.~Starobinsky, A.~Vikman and V.~Volkova for 
helpful discussions. This work has been
supported by Russian Science Foundation Grant No. 19-12-00393.

\appendix 
\section{Explicit expressions used in the second and third order calculations}
\label{appA}
In this Appendix, we give complete
expressions for Lagrangians which are
quadratic and cubic in metric perturbations in the
unitary gauge. These expressions are valid in the general Horndeski theory
whose Lagrangian is
    \begin{eqnarray}
        \mathcal{ L}&=&G_2(\phi, X)-G_3(\phi, X)\Box\phi
        +G_4(\phi, X)R+G_{4X}\left[(\Box\phi)^2-(\nabla_\mu\nabla_\nu\phi)^2\right]
        \nonumber\\&&+
        G_5(\phi, X)G_{\mu\nu}\nabla^\mu\nabla^\nu\phi
        -\frac{1}{6}G_{5X}\bigl[(\Box\phi)^3
        -3\Box\phi(\nabla_\mu\nabla_\nu\phi)^2+
        2(\nabla_\mu\nabla_\nu\phi)^3\bigr].
    \end{eqnarray}
In each of the sectors, after writing the general expressions, we specify to
our particular model    
(\ref{Hor_L}) and (\ref{adm_func_lagr}).

\subsection*{Second-order Lagrangian in the scalar sector}

We start with the second-order Lagrangian. In the scalar sector we have
    \begin{align}
        N=1+\alpha \; , \;\;\;\; 
        N_i=\partial_i\beta\; , \;\;\;\; 
        \gamma_{ij}=a^2e^{2\zeta} \delta_{ij} \; .
    \end{align}
The quadratic Lagrangian for perturbations reads \cite{Gao:2012ib}
    \begin{eqnarray}
        \mathcal{ L}_{ss}= a^3\left[
        -3\mathcal{ G}_{T}\frac{\dot\zeta^2}{N^2}+\frac{\mathcal{ F}_T}{a^2}\zeta_{,i}\zeta_{,i}
        +\Sigma \alpha^2
        -\frac{2}{a^2}\Theta\alpha\beta_{,ii}
        +\frac{2}{a^2}\mathcal{ G}_T\frac{\dot \zeta}{N}\beta_{,ii}
        +6\Theta \alpha\frac{\dot\zeta}{N}
        -\frac{2}{a^2}\mathcal{ G}_T\alpha\zeta_{,ii}
        \right],\label{scalaraction1}
    \end{eqnarray}
where 
    \begin{subequations}
    \label{full_sec_ord_tensor}
    \begin{eqnarray}
        \mathcal{ F}_T&=&2\left[G_4
        -X\left( \frac{\ddot\phi}{N^2} G_{5X}+G_{5\phi}\right)\right],
        \\
        \mathcal{ G}_T&=&2\left[G_4-2 XG_{4X}
        -X\left(H\frac{\dot\phi}{N} G_{5X} -G_{5\phi}\right)\right],
    \end{eqnarray}
    \end{subequations}
          
    \begin{subequations}
    \begin{eqnarray}
        \Sigma&=&XG_{2X}+2X^2G_{2XX}+12H\frac{\dot\phi}{N} XG_{3X}
        +6H\frac{\dot\phi}{N} X^2G_{3XX}
        -2XG_{3\phi}-2X^2G_{3\phi X}
        \nonumber\\&&
        -6H^2G_4
        +6\Bigg[H^2\left(7XG_{4X}+16X^2G_{4XX}+4X^3G_{4XXX}\right)
        \nonumber\\&&
        -H\frac{\dot\phi}{N}\left(G_{4\phi}+5XG_{4\phi X}+2X^2G_{4\phi XX}\right)
        \Bigg]
        \nonumber\\&&
        +30H^3\frac{\dot\phi}{N} XG_{5X}+26H^3\frac{\dot\phi}{N} X^2G_{5XX}
        +4H^3\frac{\dot\phi}{N} X^3G_{5XXX} 
        \nonumber\\&&
        -6H^2X\left(6G_{5\phi}
        +9XG_{5\phi X}+2 X^2G_{5\phi XX}\right),
    \end{eqnarray}

    \begin{eqnarray}
    \label{defTheta}
        \Theta&=&-\frac{\dot\phi}{N} XG_{3X}+
        2HG_4-8HXG_{4X}
        -8HX^2G_{4XX}+\frac{\dot\phi}{N} G_{4\phi}+2X\frac{\dot\phi}{N} G_{4\phi X}
        \nonumber\\&&
        -H^2\frac{\dot\phi}{N}\left(5XG_{5X}+2X^2G_{5XX}\right)
        +2HX\left(3G_{5\phi}+2XG_{5\phi X}\right).
    \end{eqnarray}
    \end{subequations}
Variation of this action with respect to the Lagrange multipliers
$\alpha$ and $\beta$
leads to the linearized constraint equations
    \begin{subequations}
    \label{constraints}
    \begin{eqnarray}
        \Sigma\alpha -\frac{\Theta}{a^2}\partial^2\beta +3\Theta\frac{\dot\zeta}{N}
        -\frac{\mathcal{ G}_T}{a^2}\partial^2\zeta&=&0,
        \\
        \Theta\alpha-\mathcal{ G}_T\frac{\dot\zeta}{N}&=&0.
    \end{eqnarray}
    \end{subequations}
The solution to  these equations is
    \begin{subequations}
    \label{a_b_constraints}
    \begin{eqnarray}
        \alpha&=&\frac{\mathcal{ G}_T}{\Theta}\frac{\dot\zeta}{N},\label{csol1}
        \\
        \beta
        &=&\frac{1}{a\mathcal{ G}_T}\left(a^3\mathcal{ G}_S\psi
        -\frac{a\mathcal{ G}_T^2}{\Theta}\zeta\right)
        \; .
        \label{csol2}
    \end{eqnarray}
    \end{subequations}
Upon inserting these expressions into \eqref{scalaraction1}, 
one finds  \cite{Kobayashi:2011nu} that the
unconstrained
second-order action  for the
scalar metric perturbation $\zeta$ is
given by Eq. (\ref{scalar2}), where the coefficients have the following form:
    \begin{subequations}
    \label{scalar2formAPP}
    \begin{eqnarray}
        \mathcal{ F}_S&=&\frac{1}{a}\frac{d}{d t}\left(\frac{a}{\Theta}\mathcal{ G}_T^2\right)
        -\mathcal{ F}_T, 
        \\
        \mathcal{ G}_S&=&\frac{\Sigma }{\Theta^2}\mathcal{ G}_T^2+3\mathcal{ G}_T. 
    \end{eqnarray}
    \end{subequations}
The sound speed is given by $c_s^2=\mathcal{ F}_S/\mathcal{ G}_S$.

For our specific Lagrangian
(\ref{Hor_L}) with $G_4=G_4(\phi)$ and $G_5(\phi,X) = 0$, 
expressions (\ref{full_sec_ord_tensor}) have the form of (\ref{tensor2form}),
while
for $\Sigma$ and $\Theta$ we have the following expressions: 
    \begin{subequations}
    \label{SigmaTheta_form}
    \begin{eqnarray}
        \Sigma&=&XG_{2X}+2X^2G_{2XX}+12H\frac{\dot\phi}{N} XG_{3X}
        +6H\frac{\dot\phi}{N} X^2G_{3XX}
        -2XG_{3\phi}-2X^2G_{3\phi X}
        \nonumber\\&&
        -6H^2G_4
        -6H\frac{\dot\phi}{N} G_{4\phi},
        \\
        \Theta&=&-\frac{\dot\phi}{N} XG_{3X}+
        2HG_4
        +\frac{\dot\phi}{N} G_{4\phi}.
    \end{eqnarray}
    \end{subequations}
In 3+1 formalism these formulas read
    \begin{subequations}
    \label{SigmaTheta_ADM}
    \begin{eqnarray}
        \Sigma&=&-3H^2\left[B_4 - \frac{1}{6H^2}\left(A_2 + 
        3N A_{2N}+N^2A_{2NN}\right)-
        \frac{1}{2H}\left(NA_{3N}+N^2A_{3NN}\right)\right],
        \\
        \Theta&=&2H\left(\frac{NA_{3N}}{4H}+B_4\right),
    \end{eqnarray}
    \end{subequations}
where we use the relation $\sqrt{2X} = \frac{1}{(-t)N}$ which is obtained 
from the gauge condition (\ref{gauge}).
The asymptotic behavior of the coefficients in the model \eqref{adm_func_lagr}
is given by Eq.~\eqref{scalar2asymp}, and we use in what follows the asymptotics
    \begin{eqnarray}
        \Sigma\propto (-t)^{-2\alpha-\delta-2}, \;\;\;\;\;\;\;
        \Theta\propto (-t)^{-2\alpha-\delta-1}.
        \label{asyTheta}
    \end{eqnarray}

\subsection*{Third-order action}
\subsubsection*{Scalar sector}
Complete third-order action for scalars $\alpha$, $\beta$ and $\zeta$ 
is given in \cite{Gao:2011qe, DeFelice:2011uc, Gao:2012ib} 
and has the following form:
    \begin{eqnarray}
        \mathcal{ L}_{sss}&=&
        -\frac{a^3}{3}\left(\Sigma+2X\Sigma_X+H\Xi\right)\alpha^3
        +a^3\left[3\Sigma\zeta+\Xi\frac{\dot\zeta}{N}+\left(\Gamma-\mathcal{ G}_T\right)
        \frac{\zeta_{,ii}}{a^2}-\frac{\Xi}{3a^2}\beta_{,ii}
        \right]\alpha^2
        \nonumber\\&&
        -2a\Theta\alpha\zeta_{,i}\beta_{,i}+18a^3\Theta\alpha\zeta\frac{\dot\zeta}{N}
        +4a\mu\alpha\frac{\dot\zeta}{N}\zeta_{,ii}
        -\frac{\Gamma}{2a}\alpha\left(\beta_{,ij}\beta_{,ij}-\beta_{,ii}\beta_{,jj}\right)
        \nonumber\\&&
        +\frac{2\mu}{a}\alpha\left(\beta_{,ij}\zeta_{,ij}-\beta_{,ii}\zeta_{,jj}\right)
        -2a\Theta\alpha\beta_{,ii}\zeta - 2a\Gamma\alpha\beta_{,ii}\frac{\dot\zeta}{N}
        -2a\mathcal{ G}_T\alpha\zeta\zeta_{,ii}
        -a\mathcal{ G}_T\alpha\zeta_{,i}\zeta_{,i}
        \nonumber\\&&
        +
        3a^3\Gamma\alpha\frac{\dot{\zeta}^2}{N^2}
        +2a^3\mu\frac{\dot{\zeta}^3}{N^3}+a\mathcal{ F}_T\zeta\zeta_{,i}\zeta_{,i }
        -9a^3\mathcal{ G}_T\frac{\dot{\zeta}^2}{N^2}\zeta
        +2a\mathcal{ G}_T\beta_{,i}\zeta_{,i}\frac{\dot\zeta}{N}
        -2a\mu\beta_{ii}\frac{\dot\zeta^2}{N^2}
        \nonumber\\&&
        +2a\mathcal{ G}_T\beta_{,ii}\frac{\dot\zeta}{N}\zeta
        +\frac{1}{a}\left(\frac{3}{2}\mathcal{ G}_T\zeta -\mu\frac{\dot\zeta}{N}\right)
        \left(\beta_{,ij}\beta_{,ij}-\beta_{,ii}\beta_{,jj}\right)
        -2\frac{\mathcal{ G}_T}{a}\beta_{,ii}\beta_{,j}\zeta_{,j},
        \label{unconstr_lagr}
    \end{eqnarray}
where
    \begin{eqnarray}
        \Xi &=&
        12\frac{\dot\phi}{N} XG_{3X}
        +6\frac{\dot\phi}{N} X^2G_{3XX}
        -12HG_4
        \nonumber\\&&
        +6\left[2H\left(7XG_{4X}+16X^2G_{4XX}+4X^3G_{4XXX}\right)
        -\frac{\dot\phi}{N}\left(G_{4\phi}+5XG_{4\phi X}+2X^2G_{4\phi XX}\right)
        \right]
        \nonumber\\&&
        +90H^2\frac{\dot\phi}{N} XG_{5X}+78H^2\frac{\dot\phi}{N} X^2G_{5XX}
        +12H^2\frac{\dot\phi}{N} X^3G_{5XXX} \nonumber\\&&
        -12HX\left(6G_{5\phi}
        +9XG_{5\phi X}+2 X^2G_{5\phi XX}\right),
    \end{eqnarray}
    \begin{eqnarray}
        \Gamma&=& 2G_4-8XG_{4X}-8X^2G_{4XX}
        \nonumber\\&&
        -2H\frac{\dot\phi}{N}\left(5XG_{5X}+2X^2G_{5XX}\right)
        +2X\left(3G_{5\phi}+2XG_{5\phi X}\right),
    \end{eqnarray}
    \begin{eqnarray}
        \mu =\frac{\dot\phi}{N} XG_{5X}.\label{mudefine}
    \end{eqnarray}
    We insert solutions   \eqref{a_b_constraints} to the
    constraints \footnote{It is worth 
noting that to obtain the unconstrained 
cubic action, it is sufficient to solve the constraint equations 
for  the Lagrange multipliers 
to the {\it first} order in perturbations.
Indeed, let 
$\alpha_i$ be the Lagrange multipliers and $\zeta_A$ be 
dynamical variables (in our case there is one such variable $\zeta$). 
The quadratic and cubic part of the 
action has the form $\mathcal{S}^{(2)}+\mathcal{S}^{(3)} = \alpha A \alpha +
\alpha B \zeta +$ $\zeta C \zeta$ $+ \mbox{cubic} (\alpha, \zeta)$,
where $A$ and $C$   are 
symmetric matrices.
To quadratic order, 
the solution to the constraint equation is $\alpha = \alpha^{(1)}
+ \alpha^{(2)}$, where the first-order term obeys
    \begin{equation}
        2A \alpha^{(1)} + B \zeta = 0
    \label{general_constr}
    \end{equation}
and $\alpha^{(2)} = O (\zeta^2)$. One inserts $\alpha = \alpha (\zeta)$
back into the original action to obtain the
unconstrained action and finds 
that the contribution of $\alpha^{(2)}$ to the unconstrained
cubic action
is $2 \alpha^{(2)} A \alpha^{(1)} + \alpha^{(2)} B \zeta$, i.e.,
it vanishes due to \eqref{general_constr}. So, the unconstrained {\it cubic}
action is obtained by plugging the {\it first-order} solution for $\alpha_i$
back 
into the original action; see Ref. \cite{Chen:2006nt}.}
    into the
Lagrangian (\ref{unconstr_lagr})
and find the unconstrained cubic \footnote{Actually, 
direct substitution of constraints (\ref{a_b_constraints}) 
into the cubic Lagrangian (\ref{unconstr_lagr}) gives 18 terms. One of them, 
namely $\Lambda_{18} \partial^2\zeta(\partial_i\psi)^2$, 
can be straightforwardly reduced to other ones by  integrating by parts.} Lagrangian for $\zeta$.
The expressions for the
coefficients of the 17 terms 
in formula (\ref{S3scalar}) 
are given by (in square brackets we also show the 
interaction type for convenience)
\renewcommand{\arraystretch}{2.05}
\[
   \begin{array}{*2{>{\displaystyle}r>{\displaystyle}l}}
        \Lambda_1 [\dot{\zeta}^3/N^3] &\multicolumn{3}{l}{\displaystyle = -\frac{\mathcal{ G}_T^3}{3\Theta^3}(\Sigma+2X\Sigma_X+H\Xi) 
        + \frac{\mathcal{ G}_T^2\Xi}{\Theta^2} 
        - \frac{\mathcal{ G}_T \mathcal{ G}_S \Xi}{3\Theta^2} 
        + \frac{\Gamma \mathcal{ G}_S^2}{2\Theta \mathcal{ G}_T} \nonumber } \\
        &\multicolumn{2}{l}{\displaystyle \phantom{=} - \frac{2\Gamma \mathcal{ G}_S}{\Theta} + \frac{3\Gamma \mathcal{ G}_T}{\Theta} 
        + \mu\Big(2-2\frac{\mathcal{G}_S}{\mathcal{G}_T}
        +\frac{\mathcal{G}_S^2}{\mathcal{G}_T^2}\Big), \nonumber } & \\
        \Lambda_2 [(\dot{\zeta}^2/N^2)\zeta] &\multicolumn{2}{l}{\displaystyle = \frac{3\mathcal{ G}_T^2 \Sigma}{\Theta^2} 
        + 9\mathcal{ G}_T - \frac{3\mathcal{ G}_S^2}{2\mathcal{ G}_T}, \nonumber } & \\
        \Lambda_3 [(\dot{\zeta}^2/N^2) \partial^2 \zeta] &\multicolumn{2}{l}{\displaystyle = \frac{\mathcal{ G}_T^3 \Xi}{3a^2\Theta^3} 
        - \frac{\mathcal{ G}_T \mathcal{ G}_S \Gamma}{a^2\Theta^2} + \frac{2\Gamma\mathcal{ G}_T^2}{a^2\Theta^2}+\frac{\mu}{a^2}\Big(6\frac{\mathcal{G}_T}{\Theta}
        -4\frac{\mathcal{G}_S}{\Theta}\Big), \nonumber } & \\
        \Lambda_4 [(\dot{\zeta}/N)\zeta \partial^2 \zeta] &=\frac{3\mathcal{ G}_T \mathcal{ G}_S}{a^2\Theta} 
        - \frac{2\mathcal{ G}_T^2}{a^2\Theta},
        &\Lambda_5 [(\dot{\zeta}/N) \left(\partial_i \zeta \right)^2] &=
        -\frac{\mathcal{ G}_T^2}{a^2\Theta} + \frac{2\mathcal{ G}_T \mathcal{ G}_S}{a^2\Theta}, \nonumber
        \\
        \Lambda_6 [\zeta \left(\partial_i \zeta \right)^2] &=\frac{\mathcal{ F}_T}{a^2},
        & \Lambda_7 [(\dot{\zeta}/N) \left(\partial^2 \zeta \right)^2] 
        &=\frac{\Gamma\mathcal{ G}_T^3}{2a^4\Theta^3}+ 3\frac{\mu\mathcal{G}_T^2}{a^4\Theta^2}, \nonumber \\
        \Lambda_8[\zeta \left(\partial^2 \zeta \right)^2]&=
        -\frac{3\mathcal{ G}_T^3}{2a^4\Theta^2},
        & \Lambda_9[\partial^2 \zeta \left(\partial_i \zeta \right)^2]&=
        -\frac{2\mathcal{ G}_T^3}{a^4\Theta^2}, \nonumber \\
        \Lambda_{10}[(\dot{\zeta}/N) \big( \partial_i \partial_j \zeta \big)^2]&=
        -\frac{\Gamma\mathcal{ G}_T^3}{2a^4\Theta^3} 
        -3\frac{\mu\mathcal{G}_T^2}{a^4\Theta^2},
        & \Lambda_{11}[\zeta \big( \partial_i \partial_j \zeta \big)^2]&=\frac{3\mathcal{ G}_T^3}{2a^4\Theta^2}, \nonumber \\
        \Lambda_{12}[(\dot{\zeta}/N) \partial_i \zeta \partial^i \psi]&=
        -\frac{2\mathcal{ G}_S^2}{\mathcal{ G}_T},
        & \Lambda_{13}[\partial^2 \zeta \partial_i \zeta \partial^i \psi]&=
        \frac{2 \mathcal{ G}_T\mathcal{ G}_S}{a^2 \Theta},\nonumber \\
        \Lambda_{14}[(\dot{\zeta}/N) \big( \partial_i \partial_j \psi \big)^2]&=
        -\frac{\Gamma\mathcal{ G}_S^2}{2\Theta\mathcal{G}_T} 
        - \frac{\mu\mathcal{G}_S^2}{\mathcal{G}_T^2},
        & \Lambda_{15}[\zeta \big( \partial_i \partial_j \psi \big)^2]&=
        \frac{3\mathcal{ G}_S^2}{2\mathcal{ G}_T},\nonumber \\
        \Lambda_{16}[(\dot{\zeta}/N) \partial_i \partial_j \zeta \partial^i \partial^j \psi]&=
        \frac{\mathcal{ G}_T\mathcal{ G}_S\Gamma}{a^2\Theta^2} 
        + 4\frac{\mu\mathcal{G}_S}{a^2\Theta},
        & \Lambda_{17}[\zeta \partial_i \partial_j \zeta \partial^i \partial^j \psi]&=
        -\frac{3\mathcal{ G}_T\mathcal{ G}_S}{a^2\Theta}.
	\end{array}
\]
\renewcommand{\arraystretch}{1}
For our specific model (\ref{Hor_L}) and \eqref{adm_func_lagr} we have 
    \begin{subequations}
    \begin{eqnarray}
        \Xi &=& 12\frac{\dot{\phi}}{N}XG_{3X} + 6\frac{\dot{\phi}}{N}X^2G_{3XX}-
        12HG_4-6\frac{\dot{\phi}}{N}G_{4\phi},  \\
        \Gamma &=& \mathcal{G}_T = \mathcal{F}_T = 2G_4,  \\
        \mu &=& 0.
    \end{eqnarray}
    \label{xi_gam_mu}
    \end{subequations}
In ADM formalism we have
    \begin{eqnarray}
        \Xi = A_3 + \frac{3}{2}N^2A_{3NN}-12HB_4,\;\;\;\Gamma = 2B_4 \; .
    \end{eqnarray}
    The asymptotic behavior of these
    combinations is    
    \begin{eqnarray}
        \Xi\propto(-t)^{-2\alpha-1-\delta},\;\;\;\Gamma \propto (-t)^{-2\alpha}.
    \end{eqnarray}
Making use of these expressions, 
we find the early-time asymptotics of 
all $\Lambda_i$, listed in Table~\ref{scalarcondtab}.

\subsubsection*{Mixed sectors}
First, let us turn to the case of two tensors and one scalar.
One has~\cite{Gao:2012ib}
\begin{subequations}
    \begin{eqnarray}
        d_1&=&\frac{3\mathcal{ G}_T}{8}\left[1-\frac{H\mathcal{ G}_T^2}{\Theta\mathcal{ F}_T}
        +\frac{\mathcal{ G}_T}{3}\frac{d}{Nd t}\left(\frac{\mathcal{ G}_T}{\Theta\mathcal{ F}_T}\right)
        \right],
        \\
        d_2&=&\frac{\mathcal{ F}_S}{8},
        \\
        d_3&=&-\frac{\mathcal{ G}_S}{4},
        \\
        d_4&=&\frac{\mathcal{ G}_T}{8\Theta \mathcal{ F}_T}\left(\mathcal{ G}_T^2-\Gamma\mathcal{ F}_T\right)
        \nonumber\\
        &+&\frac{\mu}{4}\left[
        \frac{\mathcal{ G}_S}{\mathcal{ G}_T}-1-\frac{H\mathcal{ G}_T^2}{\Theta\mathcal{ F}_T}
        \left(6+\frac{\dot{\mathcal{ G}}_S}{NH\mathcal{ G}_S}\right)\right]
        +\frac{\mathcal{ G}_T^2}{4}\frac{d}{Nd t}\left(\frac{\mu}{\Theta\mathcal{ F}_T}\right),
        \\
        d_5&=&\frac{\mu\mathcal{ G}_T}{4\Theta}\left(\frac{\mathcal{ F}_S\mathcal{ G}_T}{\mathcal{ F}_T\mathcal{ G}_S}
        -1\right) \; ,
        \\
        d_6&=&-\frac{\mu}{2}\frac{\mathcal{ G}_S}{\mathcal{ G}_T} \; ,
        \\
        d_7&=&\frac{\mu}{2}\frac{\mathcal{ G}_T}{\Theta} \; .
    \end{eqnarray}
\end{subequations}
    Several coefficients vanish due to \eqref{xi_gam_mu}:
    \begin{equation*}
        d_4= d_5 = d_6 = d_7 = 0 .
    \end{equation*}
        The asymptotics of the remaining coefficients are given in
        Table~\ref{crosscondtab}.     
      
        We now write formulas for two scalar and one tensor case~\cite{Gao:2012ib}:
        \begin{subequations}
    \begin{eqnarray}
        c_1&=&\mathcal{ F}_S,
        \\
        c_2&=&\frac{\Gamma}{4\Theta}\left(\mathcal{ F}_S-\mathcal{ F}_T\right)
        +\frac{\mathcal{ G}_T^2}{\Theta}\left[
        -\frac{1}{2}+\frac{H\Gamma}{4\Theta}\left(3
        +\frac{\dot{ \mathcal{ G}}_T}{NH\mathcal{ G}_T}\right)
        -\frac{1}{4}\frac{d}{Nd t}\left(\frac{\Gamma}{\Theta}\right)
        \right]
        \nonumber\\&&
        +\frac{\mu\mathcal{ F}_S}{\mathcal{ G}_T}+\frac{2H\mathcal{ G}_T\mu}{\Theta}
        -\mathcal{ G}_T\frac{d}{Nd  t}\left(\frac{\mu}{\Theta}\right),
        \\
        c_3&=&\mathcal{ G}_S\left[
        \frac{3}{2}+\frac{d}{Nd t}\left(\frac{\Gamma}{2\Theta}+\frac{\mu}{\mathcal{ G}_T}\right)
        -\left(3H+\frac{\dot{\mathcal{ G}}_T}{N\mathcal{ G}_T}\right)
        \left(\frac{\Gamma}{2\Theta}+\frac{\mu}{\mathcal{ G}_T}\right)
        \right],
        \\
        c_4&=&\mathcal{ G}_S\left[
        -\frac{\mathcal{ G}_T^2-\Gamma\mathcal{ F}_T}{2\Theta\mathcal{ G}_T}
        -\frac{2H\mu}{\Theta}+\frac{d}{Nd t}\left(\frac{\mu}{\Theta}\right)
        +\frac{\mu}{\mathcal{ G}_T^2}\left(\mathcal{ F}_T-\mathcal{ F}_S\right)
        \right],
        \\
        c_5&=&\frac{\mathcal{ G}_T^2}{2\Theta}\left[
        \frac{\mathcal{ G}_T^2-\Gamma\mathcal{ F}_T}{2\Theta\mathcal{ G}_T}
        +\frac{2H\mu}{\Theta}-\frac{d}{Nd t}\left(\frac{\mu}{\Theta}\right)
        -\frac{\mu}{\mathcal{ G}_T^2}\left(3\mathcal{ F}_T-\mathcal{ F}_S\right)
        \right],
        \\
        c_6&=&\frac{\mathcal{ G}_S^2}{4\mathcal{ G}_T}\left[
        1+\frac{6H\mu}{\mathcal{ G}_T}-2\mathcal{ G}_T\frac{d}{Nd t}\left(\frac{\mu}{\mathcal{ G}_T^2}\right)
        \right].
    \end{eqnarray}
    \end{subequations}
    We again make use of \eqref{xi_gam_mu} and obtain
    \begin{equation*}
      c_4=c_5=0 .
    \end{equation*}
    Other coefficients have asymptotics listed 
        in Table \ref{crosscondtab}.

\section{Einstein frame: strong coupling problem in models with singular metric }
\label{appB}

In this Appendix we point out that when dealing with singular metric,
one should keep in mind the potential 
strong coupling problem, even if it does not show up at quadratic order.
To this end, we
consider the theory  \eqref{Hor_L} and make the
conformal transformation of the metric:
    \begin{equation}
    \label{conftrans}
        \tilde{g}_{\mu\nu} = \Omega(\phi)g_{\mu\nu}, 
    \end{equation}
where $\Omega(\phi)$ is a conformal factor. 
The new metric has the standard FLRW form
$\tilde{ds^2} = -\tilde{N}^2dt^2 + \tilde{a}^2 dx^2$,
where
    \begin{eqnarray}
    \label{Einst_a_N}    
        \tilde{N}^2 = \Omega N^2, \hspace{5mm} \tilde{a}^2 = \Omega a^2,
    \end{eqnarray}
    We take $\Omega(\phi) = 2G_4(\phi)/M^2_{Pl}$ and thus move from
    the Jordan frame and the Lagrangian \eqref{Hor_L} to the
Einstein frame and the Lagrangian
    \begin{eqnarray}
        \cal \tilde{L}&=&\tilde{G}_2(\phi, \tilde{X})-\tilde{G}_3(\phi, \tilde{X})\tilde{\Box} \phi
        + \frac{M_{Pl}^2}{2}\tilde{R},
    \end{eqnarray}
with $\tilde{X}  =  -\frac{1}{2}\tilde{g}^{\mu\nu}\tilde{\partial}_{\mu}\phi\tilde{\partial}_{\nu}\phi$, $\tilde{\Box} \phi = \tilde{g}^{\mu\nu}\tilde{\nabla}_{\mu}\tilde{\nabla}_{\nu}\phi$, 
where the functions $\tilde{G}_2$, $\tilde{G}_3$ and $\tilde{G}_4$  
are combinations of the Jordan frame $G_2, G_3$ and $G_4$,
and $\tilde{R}$ is the Ricci 
scalar for the metric  $\tilde{g}_{\mu\nu}$.

Since $G_4 =B_4$ is given by 
\eqref{B_4}, it is straightforward to find the asymptotics of the
Einstein frame background solution in terms of
cosmic time  $\tau \propto \int\sqrt{\Omega}Ndt \propto -(-t)^{1-\alpha}$.
We consider the case $\alpha<1$, in which strong coupling is absent in the
tensor sector, and
get
    \begin{subequations}
    \begin{align}
        \tilde{a} &\propto  \frac{1}{(-\tau)^{\alpha/(1-\alpha)}}, \\
        \tilde{H} &\propto \frac{1}{(-\tau)}, \\ e^{\phi}&\propto \frac{1}{(-\tau)^{1/(1-\alpha)}}\; .
    \end{align}
    \end{subequations}
    The cosmological evolution starts at $\tau \to -\infty$. Notably, the metric
    is singular in this asymptotics,
    $\tilde{a}\to0$ as $\tau\to -\infty$, but the Hubble parameter
    and its
    derivatives vanish. A similar
solution was called ``modified genesis'' 
in Ref.~\cite{Libanov:2016kfc}, where it was assumed that such solutions are
healthy. The analysis of this paper shows, however, that
this is not necessarily the case: away from the solid triangle and above
the blue line
in
Fig.~\ref{possible_param}, our model suffers from the strong coupling problem.
Indeed, the range of parameters in which the model is healthy (not healthy)
 is invariant under field redefinition
(conformal transformation in our case).

To emphasize the subtlety of the situation,
let us consider linearized theory in the Einstein frame. The metric is
    \begin{equation}
        d\tilde{s}^2=-\tilde{N}^2 dt^2+\tilde{\gamma}_{ij}\left( dx^i+\tilde{N}^i dt\right)\left(dx^j+\tilde{N}^j dt\right),
    \end{equation}
where 
    \begin{equation}
      \tilde{\gamma}_{ij}=\tilde{a}^2e^{2\zeta} \left(e^{h}\right)_{ij}, \;\;\; \delta \tilde{N}= \tilde{N} \tilde{\alpha},
      \;\;\; \delta \tilde{N}_i=\partial_i\tilde{\beta}
    \end{equation}
and thus  
    \begin{equation}
        \tilde{\alpha} = \alpha, \;\;\; \tilde{\beta} = \beta/\sqrt{\Omega}, \;\;\; \tilde{\zeta} = \zeta.
    \end{equation}
    The scalar perturbation $\tilde{\zeta}=\zeta$ has the same action
    as in the Jordan frame, so in terms of the Einstein frame
    variables one has
       \begin{eqnarray}
        \tilde{\mathcal{ S}}_{ss}=\int \tilde{N}dt \tilde{a}^3d^3x\left[
        \tilde{\mathcal{ G}}_S
        \frac{\dot\zeta^2}{\tilde{N}^2}
        -\frac{\tilde{\mathcal{ F}}_S}{\tilde{a}^2}
        \zeta_{,i}\zeta_{,i}
        \right]\label{scalar2_einst},
    \end{eqnarray}
where
    \begin{equation}
        \tilde{\mathcal{G}}_S = \frac{\mathcal{G}_S}{\Omega}, \;\;\;  
        \tilde{\mathcal{F}}_S = \frac{\mathcal{F}_S}{\Omega}.
    \end{equation}
The asymptotic behavior of the latter coefficients is 
    \begin{equation}
        \tilde{\mathcal{G}}_S \propto (-\tau)^{\delta/1-\alpha}, \;\;\;  
        \tilde{\mathcal{F}}_S \propto (-\tau)^{\delta/1-\alpha} \; .
    \end{equation}
Unlike in the Jordan frame, 
these coefficients tend  to $\infty$ as $\tau \to -\infty$.
Naively, this behavior suggests that not only is there no danger of strong
coupling, but the theory becomes free in the asymptotic past.
Were the background metric nonsingular, this would indeed be the case. In
our model with a singular Einstein frame metric, the naive expectation fails:
the model is strongly coupled as $\tau \to -\infty$ (away from
 the solid triangle and above the blue line in
 Fig.~\ref{possible_param}). We conclude that in models with a singular metric,
 the study of the quadratic action for perturbations is insufficient for
 analyzing the strong coupling problem.


\end{document}